\documentclass[conference]{IEEEtran}

\IEEEoverridecommandlockouts
\usepackage{cite}
\usepackage{amsmath,amssymb,amsfonts}
\usepackage{algorithmic}
\usepackage{graphicx}
\usepackage{textcomp}
\usepackage{xcolor}

\usepackage{booktabs}
\usepackage{makecell}
\usepackage{tabularx}
\usepackage{hyperref}
\usepackage{authblk}

\def\BibTeX{{\rm B\kern-.05em{\sc i\kern-.025em b}\kern-.08em
    T\kern-.1667em\lower.7ex\hbox{E}\kern-.125emX}}
    
\begin{document}

\title{Securing RAG:\\ A Risk Assessment and Mitigation Framework}

\author[1,**]{Lukas Ammann}
\author[1,**]{Sara Ott}
\author[1,2]{Christoph R. Landolt}
\author[1,*]{Marco P. Lehmann}

\affil[1]{Eastern Switzerland University of Applied Sciences (OST), Rapperswil, Switzerland} 
\affil[2]{Cyber-Defence Campus, armasuisse Science and Technology, Thun, Switzerland}
\affil[*]{Corresponding author: marco.lehmann@ost.ch}
\affil[**]{Equal contribution}

% \author{\IEEEauthorblockN{Lukas Ammann}
% \IEEEauthorblockA{\textit{School of Computer Science } \\
% \textit{Eastern Switzerland University of Applied Sciences} (OST)\\
% Rapperswil, Switzerland  \\
% \url{https://orcid.org/0009-0006-5149-2342}}
% \and
% \IEEEauthorblockN{Sara Ott}
% \IEEEauthorblockA{\textit{School of Computer Science } \\
% \textit{Eastern Switzerland University of Applied Sciences} (OST)\\
% Rapperswil, Switzerland  \\
% \url{https://orcid.org/0009-0003-3950-9736}}
% \and
% \IEEEauthorblockN{Christoph R. Landolt}
% \IEEEauthorblockA{\textit{School of Engineering} \\
% \textit{Eastern Switzerland University of Applied Sciences} (OST)\\
% Rapperswil, Switzerland  \\
% \url{https://orcid.org/0009-0002-7031-3291}}
% \and
% \IEEEauthorblockN{Marco Lehmann}
% \IEEEauthorblockA{\textit{School of Computer Science} \\
% \textit{Eastern Switzerland University of Applied Sciences} (OST)\\
% Rapperswil, Switzerland  \\
% \url{https://orcid.org/0000-0001-5274-144X}}
% }

\maketitle

% Todo: verify after acceptance/publication:
% https://conferences.ieeeauthorcenter.ieee.org/author-ethics/guidelines-and-policies/post-publication-policies/#preprint

\textit{This work has been submitted to the IEEE for possible publication. Copyright may be transferred without notice, after which this version may no longer be accessible.} \\

\begin{abstract}

Retrieval Augmented Generation (RAG) has emerged as the de facto industry standard for user-facing NLP applications, offering the ability to integrate data without re-training or fine-tuning Large Language Models (LLMs). 
This capability enhances the quality and accuracy of responses but also introduces novel security and privacy challenges, particularly when sensitive data is integrated. 
With the rapid adoption of RAG, securing data and services has become a critical priority.
This paper first reviews the vulnerabilities of RAG pipelines, and outlines the attack surface from data pre-processing and data storage management to integration with LLMs.
The identified risks are then paired with corresponding mitigations in a structured overview. 
In a second step, the paper develops a framework that combines RAG-specific security considerations, with existing general security guidelines, industry standards, and best practices. 
The proposed framework aims to guide the implementation of robust, compliant, secure, and trustworthy RAG systems.

\end{abstract}


\section{Introduction}

Large Language Models (LLMs) are powerful tools, leveraging extensive parametric memory to store knowledge acquired during training. However, their parametric memory inherently limits them to the data available at the time of training. 
Consequently, LLMs lack knowledge about events that occurred after training or facts that may be hidden in non-public or proprietary data, resulting in inaccuracies or hallucinations in their outputs \cite{ji-etal-2023-towards, huang2023survey}.
To address these limitations, Lewis et al. \cite{lewis2021retrievalaugmentedgenerationknowledgeintensivenlp} introduced RAG, a framework that combines the strengths of pre-trained LLMs with non-parametric memory, enabling the integration of external documents and providing a mechanism for knowledge updates. The enhanced versatility of RAG systems, combined with their improvements over standard LLMs, has opened up new use cases and led to widespread interest in business applications \cite{zhao2024retrievalaugmentedgenerationaigeneratedcontent, gao2024retrievalaugmentedgenerationlargelanguage, schillaci2024llmadoption}.
However, RAG systems introduce new security risks, such as the potential leakage of sensitive data or manipulating behavior in user-facing applications \cite{gupta2023chatgpt, zeng2024goodbadexploringprivacy}. With the widespread adoption of RAG applications, the reputational and financial risks become significant, making the protection of RAG systems mission-critical.

Given the recency of the field, the systematic analysis of RAG-related security issues lags behind the rapid evolution and adoption of the technology. 
Current discussions of attacks and mitigations are often presented informally, for example in blog posts, industry reports, or case studies. 
This scattering of information impedes the implementation of responsible and secure systems.
However, as RAG continues to gain traction, security analysts and engineers would benefit from a quick and structured guide to address emerging security challenges.
To address this need, we take a twofold approach: 
First, we conduct a comprehensive literature review to identify emerging attack vectors and corresponding mitigation strategies. 
Second, we propose a structured framework that incorporates RAG-specific security measures within a broader context: Securing a RAG System is not an isolated task but requires a holistic approach to assess system risks, balance mitigations with available resources, and  ensures regulatory compliance \cite{haryanto2024secgenaienhancingsecuritycloudbased}.
From a security perspective, RAG systems inherit the risks of the underlying system and expand the attack surface \cite{gupta2023chatgpt, kucharavy2024llmcybersecuritybook, owasp_llm_top10}. 
Furthermore, the growing system complexity and the fast development of the LLM/RAG applications make it challenging for organizations to keep up their responsible AI (RAI) programs \cite{mitsloan2023, wefcybersecurity2025}.

The paper is structured as follows:
\begin{itemize}
    \item We first describe a standard RAG architecture to establish a reference for analyzing the attack surface (Fig \ref{fig:pipeline}).
    \item We then reviewed scientific literature and industry reports for known RAG risks and mitigation strategies. A condensed summary is presented in section \ref{section:RiskMitigations}. We analyzed and structured the fragmented literature into the \textit{risk and mitigation matrix} (Table \ref{tab:risk_mitigation_table}).    
    % \item In section \ref{section:RiskMitigations}, we review literature and provide a condensed summary of risks and mitigation strategies. We then synthesize existing but fragmented literature into the \textit{risk and mitigation matrix} (Table \ref{tab:risk_mitigation_table}).
    % \item 
    Finally, in section \ref{section:Guidelines}, \textit{Guidelines for Implementation}, we integrate our RAG-specific analysis into the broader context of general security practices, and compiled a high-level framework (Fig. \ref{fig:security-layers}) for quick orientation. 
\end{itemize}

% From the review:
% Our contribution is threefold:
% a) We integrate a broad range of vulnerabilities into a coherent visualization of the attack surface (Figure 1), which also serves as a quick reference index for different risks.
% b) We systematically structure existing but fragmented information into a risk and mitigation matrix (Table 1).
% c) We contextualize RAG-specific risks and mitigations within general security standards, promoting a holistic approach to security (Figure 2).
% We are convinced that these contributions will be of great interest to the reader.

\section{Background and Related Work}
The security of machine learning systems has garnered significant attention due to increasing risks as well as the influence of legislative regulations (e.g. EU AI Act  \cite{eu_ai_regulatory_framework} and the US Executive Order on AI \cite{biden_executive_order_ai_2023}). These initiatives emphasize the need for robust instruments to ensure security, transparency, and fairness in AI systems. Complementary to these legislative efforts are technical frameworks like the NIST AI Risk Management Framework (AI RMF) \cite{nist_ai_rmf} and the \textit{Open Worldwide Application Security Project} (OWASP) Top Ten for LLM Applications \cite{owasp_llm_top10}, which address specific vulnerabilities in Large Language models (LLMs).
RAG systems have introduced new challenges to the AI threat landscape. Extensive reviews, such as \cite{zhao2024retrievalaugmentedgenerationaigeneratedcontent, gao2024retrievalaugmentedgenerationlargelanguage}, highlight the importance of securing these systems. The \emph{OWASP 2025 Top 10 Risks for LLMs and Gen AI Apps} \cite{owasp_llm_top10} include specific risks associated with RAG systems, including prompt injection \cite{greshake2023youvesignedforcompromising}, data leakage \cite{zeng2024goodbadexploringprivacy}, embedding inversion \cite{huang2024transferableembeddinginversionattack}, data poisoning \cite{zou2024poisonedragknowledgecorruptionattacks}, cross-context information conflicts \cite{xu2024knowledgeconflictsllmssurvey}, and unintentional behavior alterations \cite{zhang2024hijackraghijackingattacksretrievalaugmented}, which compromise model outputs, expose sensitive information, and undermine trustworthiness and integrity.

\subsection{Background on RAG Systems}
\begin{figure}[h]
    \centerline{\includegraphics[width=0.5\textwidth]{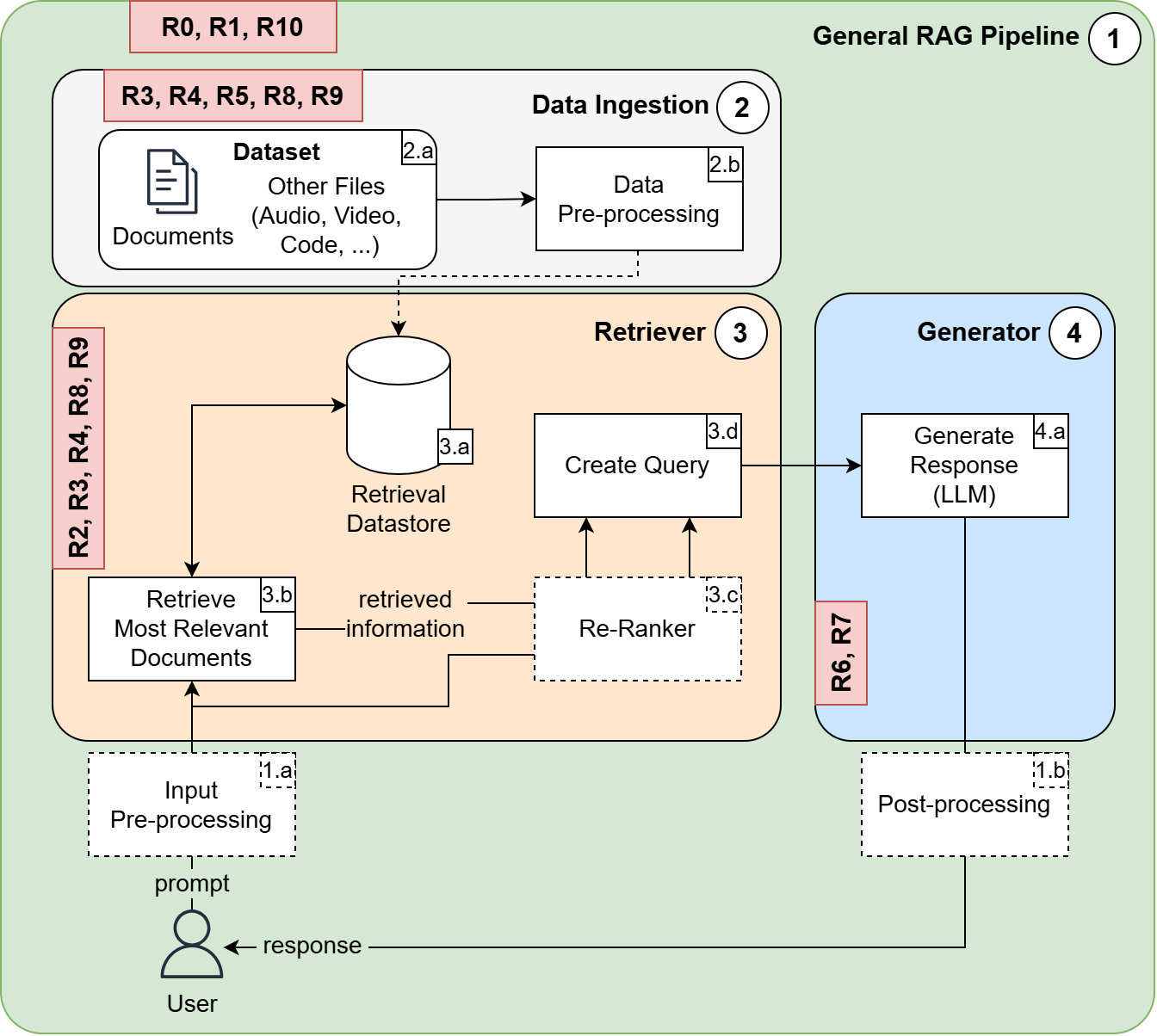}}
    \caption{General RAG Pipeline (1): A general architecture with the main components  Data Ingestion (2), Retriever (3), and Generator LLM (4). R0 to R10 indicate the risks associated with each component.}
    \label{fig:pipeline}
\end{figure}

To effectively analyze vulnerabilities and their mitigations in RAG Systems, it is essential to understand their architecture. 
This section describes a generic RAG architecture, outlining a high-level, abstract representation of its core components. This helps in understanding the attack surface and identifying potential security risks.\\
The four main components are shown in Figure \ref{fig:pipeline} and briefly discussed here:

\textbf{1. General RAG Pipeline:} 
This component represents the entire system, encapsulating all processes from user input to response generation. 
The risks in this box  do not target a specific component, but affect the user or peripheral systems.
The pipeline includes the user interface, input mechanisms, and interactions between components. 
While optional, practical RAG pipelines often include \textit{a. Pre-processing} and \textit{b. Post-processing} components, typically used for input validation and output evaluation.
\textbf{2. Data Ingestion:} This process involves asynchronous (offline) preprocessing and adding new data to the system. It consists of two subcomponents:  
\textit{a. Dataset:} Contains the information, such as documents, that the system requires to provide accurate answers using the language model.  
\textit{b. Data Pre-processing:} Transforms data into a format suitable for retrieval by chunking documents and calculating embedding vectors.
\textbf{3. Retriever:} This component identifies the most relevant documents aligned with the user’s query and passes them to the generator. It includes:  
\textit{a. Retrieval Datastore:} Optimized storage for datasets, which may include a vector database but does not strictly require one.  
\textit{b. Retrieve Documents:} Searches for the most relevant documents (or chunks) using similarity measures between vectorized prompts and documents.  
\textit{c. Re-Ranker:} An additional LLM can be used to analyse the semantics of the retrieved documents and re-rank them against the user prompt.
\textit{d. Create Query:} Combines the user prompt with retrieved information to form an augmented query, potentially including additional instructions for the generator.

\textbf{4. Generator:} A pre-trained LLM generates the answer based on the augmented query.

\section{RAG Risks and Mitigation Strategies} \label{section:RiskMitigations}
\subsection{Methods}
We identified risks and mitigations through a comprehensive literature review. 
To capture recent developments and trends, we also incorporated non-peer-reviewed sources, including preprints, blog posts, and industry reports. While this may reduce the technical depth of some findings, it provides a broader understanding of the challenges currently faced by industries adopting RAG. 

Instead of reiterating the specific technical details of the attacks, which are addressed in the cited references, we compiled a concise overview of the vulnerabilities associated with RAG adoption.
While the list presented here cannot be exhaustive, it aims to cover the RAG attack surface through a set of representative vulnerabilities.
In Figure \ref{fig:pipeline}, the risks R0 to R10 are associated with the targeted components, while Table \ref{tab:risk_mitigation_table} maps the identified risks to their corresponding mitigations. 
They are grouped into categories to improve the framework's organization and facilitating their application.

\begin{table*}[htb!]
\caption{Risk and Mitigation Matrix}\label{tab:risk_mitigation_table}
\newcommand*\rot[1]{\hbox to1em{\hss\rotatebox[origin=br]{-90}{#1}}}
\begin{tabular}{l|cccc|ccc|c|ccccc|}
\multicolumn{1}{c|}{} & 
\multicolumn{4}{c|}{\begin{tabular}[c]{@{}c@{}}Secure \\ Retrieval\end{tabular}} & 
\multicolumn{3}{c|}{\begin{tabular}[c]{@{}c@{}}System \\ Hardening\end{tabular}} & 
\multicolumn{1}{c|}{\begin{tabular}[c]{@{}c@{}}Disclosure \\ Prevention\end{tabular}} & 
\multicolumn{5}{c|}{\begin{tabular}[c]{@{}c@{}}Other \\ Mitigations\end{tabular}} \\ 
                                                 
& \rot{M0: Anonymization} & \rot{M1: Pseudonymization} & \rot{M2: Synthetic Data} & \rot{M3: Access Limitation} & \rot{M4: System Instructions} & \rot{M5: Input Validation} & \rot{M6: Evaluation} & \rot{M7: Self-hosted AI-models}  & \rot{M8: Adding Noise} & \rot{M9: Distance Threshold} & \rot{M10: Summarization} & \rot{M11: Re-Ranking} & \rot{M12: Exposure Minimization}  \\ \hline
General && & & & & & & & & & & & \\
+ R0: Blind spot and zero-day vulnerability    & x & x & x & x &   & x &   &   &   &   &   &  & x  \\
+ R1: System Limitation                    &   &   &   &   &   &   & x    &   &   &   &   &  &   \\ \hline
Vectorization and Retrieval && & & & & & & & & & & & \\
+ R2: Retrieval Data Leakage                       & x &   & x & x & x & x & x &   & x & x & x & x & x \\
+ R3: Embedding Inversion Attack                   & x & x & x &   &   &   &   & x & x &   &   &  &   \\
+ R4: Membership Inference Attack (MIA)            & x &   & x & x & x & x & x &   & x & x & x &  & x   \\
+ R5: Retrieval Data Disclosure during embedding   & x & x & x &   &   &   &   & x &   &   &   &  & x  \\ \hline
Operational Risks & & & & & & & & & & & & & \\
+ R6: Retrieval Data Disclosure during prompting    & x & x & x &   &   & x &   & x & x & x & x & x & x \\
+ R7: Prompt Disclosure                             & x & x &   &   &   & x &   & x &   &   &   &  &   \\ \hline
Data Manipulation & & & & & & & & & & & & & \\
+ R8: Knowledge Corruption Attack                  &   &   &   & x & x & x & x &   &   & x & x & x &  \\
+ R9: Indirect Prompt Injection                    &   &   &   & x & x & x & x &   &   & x & x & x &  \\
+ R10: Indirect Jailbreak Attack                   &   &   &   & x & x & x & x &   &   & x & x & x &  \\ 
\end{tabular}
\end{table*}

\subsection{Risks}

\textbf{R0: Blind spot and zero-day vulnerability} pose a challenge for RAG systems, as not all risks can be identified, and patches may not fix all problems. 
Zero-day vulnerabilities arise from factors that are currently unknown or undisclosed\cite{ablon2017zero}. This emphasizes the need for proactive risk management and continuous monitoring.

\textbf{R1: System Limitation} comprises the system inherent shortcomings of RAGs, in particular in handling and retrieving complex, multi-dimensional data. Retrieved documents may not fully address prompts, and generators can struggle to extract correct information from large input contexts, compromising system integrity.
While parametric memory evaluation is well-studied \cite{chang2023surveyevaluationlargelanguage, guo2023evaluatinglargelanguagemodels}, non-parametric memory presents unique challenges and requires a critical evaluation of the output.

\textbf{R2: Retrieval Data Leakage} refers to an attack in which information from the retrieval dataset is exposed to unauthorized entities \cite{zeng2024goodbadexploringprivacy}. This dataset may contain sensitive information, ranging from Personally Identifiable Information (PII) to more general confidential domain-specific data. Two primary categories of attack can be distinguished: In a \emph{targeted attack}, specific information is extracted from the retrieval dataset. In an \emph{opportunistic attack}, the attackers collect large volumes of data and then sift through it later to identify valuable information.

\textbf{R3: Embedding Inversion Attack} aims to reconstruct the original data from the embeddings. 
The specific methods used in such an attack can vary significantly. 
Some approaches leverage embeddings in conjunction with the exact embedding model used for their computation \cite{morris2023textembeddingsrevealalmost}. 
Other strategies involve training a surrogate model to approximate the embedding model using the original data and the associated embeddings \cite{huang2024transferableembeddinginversionattack}.
The ability to reconstruct original data with alarmingly high accuracy highlights a critical security risk. 
Consequently, embeddings require the same level of protection as the original data.
\cite{morris2023textembeddingsrevealalmost, huang2024transferableembeddinginversionattack}.

\textbf{R4: Membership Inference Attack (MIA)} attempts to determine whether specific data were included in the retrieval dataset used to populate the retrieval datastore \cite{anderson2024dataretrievaldatabasemembership}. This attack is categorized into two types: In \emph{sample-level inference}, the attacker aims to ascertain whether a specific sample x was used to populate the datastore. 
In \emph{user-level inference}, the goal is to determine if data associated with a particular user was used. 

\textbf{R5: Retrieval Data Disclosure during embedding} refers to the unintended exposure of data from the retrieval datastore to unauthorized individuals during the embedding process \cite{zeng2024goodbadexploringprivacy}. 
As the entire dataset passes through the embedding step, a successful attack may have severe consequences. 
The risk increases further if the vectorization is carried out by an external service provider.

\textbf{R6: Retrieval Data Disclosure during prompting} refers to the unintended exposure of data from the retrieval datastore to unauthorized individuals during the prompting process. This risk emerges as the most relevant documents from the retrieval datastore are retrieved and transmitted to the generator. Although this step involves only a subset of the dataset, it remains a critical concern, especially when sensitive data is involved, as prompts can be intentionally crafted to target specific documents \cite{yang2023review, jiang2024ragthief}.

\textbf{R7: Prompt Disclosure} refers to the privacy risk arising when sensitive information is included in a prompt transmitted to a cloud-hosted LLM. This risk is analogous to the established concern of search engines monitoring and recording user-entered queries \cite{duan2024privacyriskincontextlearning, chu2024reconstructpreviousconversationscomprehensively}. 
While this phenomenon may seem self-evident, it poses significant privacy concerns due to the potential exposure of sensitive data. The risk is further exacerbated in the context of chatbots, where users may be tempted to copy and paste extensive prompts, inadvertently including confidential information. 

\textbf{R8: Knowledge Corruption Attack} aims to compromise the responses of RAG systems  by manipulating the retrieval process \cite{zou2024poisonedragknowledgecorruptionattacks},  \cite{chen2024blackboxopinionmanipulationattacks}. These attacks work by influencing the retrieval of specific or manipulated documents to skew the system’s outputs.
One common approach involves altering retrieval ranking results \cite{chen2024blackboxopinionmanipulationattacks}. This is achieved by injecting manipulated text into documents that align with a specific opinion, thereby artificially increasing their relevance. This increases the likelihood that these manipulated documents will rank highly and be passed to the generator.
There are multiple methods for injecting manipulated documents into a RAG system. Attackers may directly alter the retrieval datastore by adding, modifying, or removing documents. Alternatively, if the retrieval system itself scrapes data, the documents in the specific sources can be modified \cite{zou2024poisonedragknowledgecorruptionattacks}.

\textbf{R9: Indirect Prompt Injection} refers to a security risk where malicious prompts embedded within retrieved documents manipulate the generator to execute unwanted behavior, unintended actions or produce harmful outputs \cite{owasp_llm_top10}. This risk becomes particularly critical when the generator has the capability to perform far-reaching actions, such as executing API calls or interacting with action modules to interact with external systems.

\textbf{R10: Indirect Jailbreak Attack} is designed to bypass security protocols and manipulate the generator into generating malicious or harmful content. Jailbreak attacks can be broadly categorized into two types: direct and indirect. Direct jailbreak attacks attempt to bypassing protection using customized prompts. However, their effectiveness has diminished with the implementation of advanced filtering techniques in modern LLMs \cite{wang2024poisonedlangchainjailbreakllms}.
Indirect jailbreak attacks do not manipulate the prompt, but work via retrieved documents to manipulate the generator indirectly. 
This method significantly increases both the success rates and the associated risks of such attacks \cite{deng2024pandorajailbreakgptsretrieval}.

\subsection{Mitigations}

\textbf{M0: Anonymization} is the process of irreversibly removing personal information from data \cite{stam2020data}.
This mitigation strategy is distinguished by two key characteristics. First, it can be employed as a preprocessing technique. Second, by ensuring that sensitive information is removed before populating the retrieval datastore, the risk of data leakage is significantly reduced. However, the removed information can no longer be included in the system's responses. Thus, anonymization requires careful consideration to balance privacy concerns against the need to include specific information for answering queries in the given use case. 
Anonymization is a non-trivial task, requiring meticulous planning and execution \cite{stam2020data}.

\textbf{M1: Pseudonymization} is a reversible process of substituting all sensitive information with pseudonyms \cite{stam2020data}.
The technique is particularly beneficial for RAG systems where sensitive information is needed to produce accurate and relevant responses.
The pseudonymization process consists of three straightforward steps: 
1: Replace identifiable data with placeholders while maintaining a mapping between the original data and its placeholders. 
2: Generate the response using the pseudonymized data.
3: Replace all placeholders in the response with the original data based on the mapping.
This approach helps to prevent sensitive data from being disclosed to unauthorized parties, while preserving sensitive information in responses. Pseudonymization can be applied at various points in the RAG pipeline, including the prompt (Input Pre-processing), documents (Data Pre-processing), or retrieved documents. 

\textbf{M2: Synthetic Data} are artificially generated items that replicate the characteristics and structure of the original data \cite{raghunathan2021synthetic}. This replication is achieved through a model trained on the original data to learn its properties. By replacing sensitive data with synthetic equivalents, RAG systems can operate with reduced risk of exposing confidential information while maintaining functionality and accuracy. 
Synthethic data is used when the RAG serves a downstream task. Development and testing can take place without disclosing real data.
For certain use cases, the synthetic data offers a viable solution to mitigate this risk while preserving system performance \cite{zeng2024mitigatingprivacyissuesretrievalaugmented}. 
The generation of non-leaking synthetic data is not a trivial task and needs careful validation.

\textbf{M3: Access Limitation} restricts a user's right to access only certain documents or services.
It is an effective strategy to mitigate several risks associated with RAG systems. 
However, implementing this approach requires an organization to establish a robust user authentication system that defines and enforces the necessary permissions for each user and document \cite{namer2024retrieval}. 
Access Limitation enables an organization to control access to documents in the retrieval datastore based on user roles, clearance levels, or other specific permissions. 
An organization may also restrict access to the service to a specific subgroup of users; the risk exposure is fundamentally different for an internal tool compared to a public, client-facing chatbot.
This strategy ensures that sensitive data are only accessible to authorized users, significantly reducing the risk of data exposure.

\textbf{M4: Reinforcement of System Instruction} refers to the enhancement of the generator specifications to ensure a more restrictive handling of the prompt and the data retrieval.
System instructions typically function as a template that integrates the user prompt, the retrieved data, and administrator-defined constraints or directives before passing them to the generator. 
Since this approach is highly dependent on the specific use case, there is no universal template that provides optimal risk mitigation across all scenarios. Therefore, suitable instructions must be designed and tailored individually for each application. As a starting point, templates relevant to their domain or area of interest can be explored \cite{xu2025mixtureofinstructionsaligninglargelanguage}.
Investing time and effort in designing, implementing, and testing robust system instructions is strongly recommended, as this mitigation strategy can address multiple risks effectively. 

\textbf{M5: Input Validation} safeguards the RAG pipeline by detecting and rejecting user prompts or documents containing malicious content or instructions \cite{yang2023review}. 
The system can also be extended to remove only unwanted parts of the prompt, such as sensitive information. 
Implemented as part of the input pre-processing component, it prevents harmful inputs from affecting subsequent steps or outputs. While effective in mitigating risks, implementing input validation is challenging due to the complexity of natural language and the diverse ways malicious content can be expressed.

\textbf{M6: Evaluation} improve the factual accuracy and reliability of RAG systems by identifying and (self-) correcting inaccuracies. This iterative process enhances output precision and mitigates risks \cite{asai2023selfraglearningretrievegenerate}.
Advanced approaches like Graph RAG \cite{edge2024localglobalgraphrag} and Multi-Head RAG \cite{besta2024multiheadragsolvingmultiaspect} address retrieval challenges by leveraging graph structures and multi-head attention for improved performance.
Non-parametric memory evaluation uses tasks like \emph{Needle In A Haystack}, testing retrieval of specific text from large contexts, as seen in Gemini \cite{geminiteam2024gemini}. Metrics such as Character-MAP \cite{kamps2007Eval}, BLEU \cite{papineni-etal-2002-bleu}, ROUGE \cite{lin2004rouge}, and BERTScore \cite{zhang2020bertscoreevaluatingtextgeneration} evaluate the textual overlap or semantic similarity. Recent evaluation frameworks like EXAM \cite{sander2021exam}, RAGAS \cite{ragas2023}, and ARES \cite{saadfalcon2024aresautomatedevaluationframework} assess retrieval relevance and generation quality, with ARES offering a standardized evaluation approach across diverse tasks.
More broadly, evaluation may also include MLOps measures such as continuous service monitoring.

\textbf{M7: Self-hosting of AI-models} has the advantage of having control over data flow, and allows secure, local execution of computations without offsite data transmission. Self-hosting may encompass the hosting of the generator LLM, the embedding service, and the additional LLMs used for evaluation and re-ranking. This approach may demand significant computational resources.

\textbf{M8: Adding Noise} at various stages of the RAG pipeline can help mitigate privacy risks and, in some cases, improve response accuracy \cite{zeng2024goodbadexploringprivacy, morris2023textembeddingsrevealalmost, powerOfNoiseRAG}. 
Techniques include adjusting embeddings \cite{zeng2024goodbadexploringprivacy}, inserting random characters \cite{morris2023textembeddingsrevealalmost}, or adding entire documents \cite{powerOfNoiseRAG}. 
Although it has been demonstrated that noise does not necessarily enhance RAG protection to a significant degree, its careful application can still yield positive results by increasing response accuracy \cite{powerOfNoiseRAG}.

\textbf{M9: Distance Threshold} is technique for controlling the data selected from the retrieval store. 
The relevance of a document relative to the user prompt is typically measured using a similarity function (for example cosine similarity).
Developers can define an appropriate threshold for the similarity score. 
Instead of retrieving the \textit{k-nearest} chunks, only \textit{up to k} sufficiently close chunks are retrieved. 
By limiting data retrieval, this method reduces the amount of exposed data.
The method has the additional (often primary) effect of reducing the presence of irrelevant or misleading data. This helps improving the accuracy of a service \cite{zeng2024goodbadexploringprivacy}. 

\textbf{M10: Summarization} serves as a mitigation strategy by aggregating retrieved information while excluding details unrelated to the user prompt, thus reducing data exposure and minimizing sensitive content \cite{zeng2024goodbadexploringprivacy}. A distinct LLM performs the summarization, creating a concise summary based on the user prompt and retrieved information. The summary then replaces the retrieved data and is passed to the generator for response generation.

\textbf{M11: Re-Ranking} uses an LLM (the Re-Ranker) to identify and filter the most relevant documents based on their relevance to the user prompt, preventing irrelevant data from advancing in the RAG pipeline. This approach reduces the data volume exposed to potential misuse while enhancing the relevance and quality of generated responses \cite{zeng2024goodbadexploringprivacy, rerankerSchemeNLU, alessio2024improving}. Although effective in mitigating privacy risks by focusing on relevant information, re-ranking alone is insufficient for comprehensive protection \cite{zeng2024goodbadexploringprivacy}; additional mitigation strategies should be considered.

\textbf{M12: Minimization of Exposure} is the counterpart to unknown risks (R0). 
Risks that are not known are mitigated by minimizing data collection, limiting service exposure, restricting access rights, and reducing complexity. 
More specifically, \textbf{Data minimization} is a universal principle, and part of many regulations, such as the GDPR \cite{euGDPR}. 
Mitigation M12 may involve developing a Data Strategy to identify critical data and services, and implementing Data Governance policies.

\section{Guidelines for Implementation}\label{section:Guidelines}

\begin{figure}[htb!]
    \centering
    \includegraphics[width=1.0\linewidth]{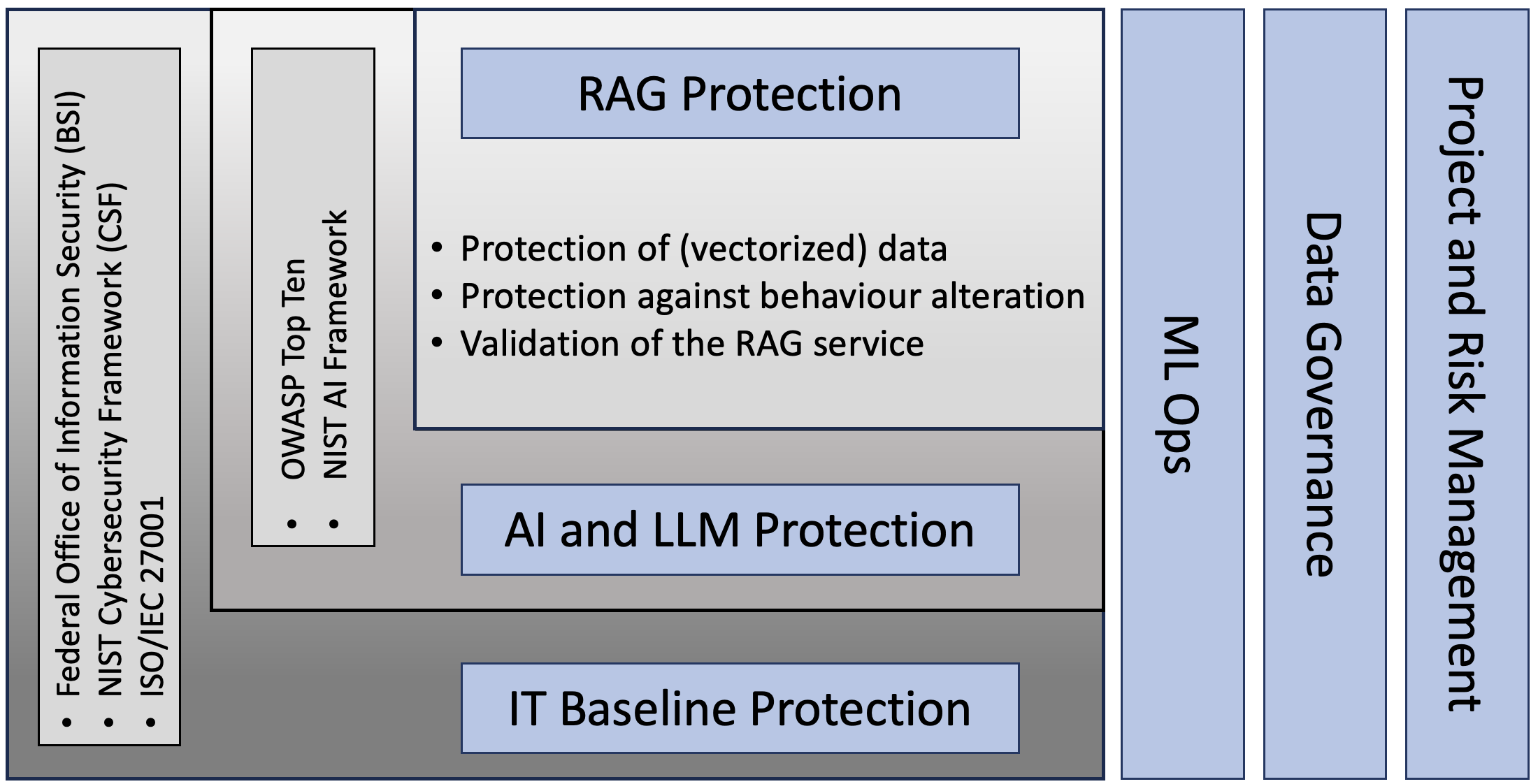}
    \caption{Securing a RAG requires a holistic approach. Three overarching activities are \emph{ML-Ops}, \emph{Data Governance}, and \emph{Project- and Risk Management}. Risks and Mitigations are addressed at layers \emph{IT Baseline Protection}, \emph{AI and LLM Protection}, and \emph{RAG Protection}.}
    \label{fig:security-layers}
\end{figure}

To implement a secure and trusted RAG, it is essential to address all organizational and technical aspects of the system, for which we propose a holistic framework (Figure \ref{fig:security-layers}) that integrates the RAG-specific Risks and Mitigations with the different standards, methodologies, and best practices of the field.

Our framework integrates three overarching activities, \emph{ML-Ops}, \emph{Data Governance}, and \emph{Project- and Risk Management}, and three security-specific layers \emph{IT Baseline Protection}, \emph{AI and LLM Protection}, and \emph{RAG Protection}.
In this section we briefly discuss these components and provide references to industry standards for practitioners.

The three overarching activities provide the ground for successful and responsible AI projects.

\textbf{Project- and Risk Management} face the additional challenges of ML and RAG projects, namely the increased system complexity, and the risks of data driven applications. 
The \emph{NIST Risk Management Framework (RMF)} \cite{NIST2024RiskManagementFramework} provides guidelines and processes to manage information security and privacy risk.
Depending on the project scope, compliance with regulations (e.g. EU AI Act \cite{euAIact} and the US Executive Order on AI \cite{biden_executive_order_ai_2023}) can be a formal requirement which requires the Project Management to integrate the necessary steps early on.

\textbf{Data Governance} is based on design principles and is a foundation of trustworthy AI \cite{janssen2020data}. It comprises policies and responsibilities for managing data, and supports safe and compliant (e.g. with GDPR) use of data \cite{blivznak2024systematic}. 
In addition, Data Governance can address more general, societal risks related to the use of AI for decision making. 

\textbf{ML-Ops} extends Continuous Integration (CI) and Continuous Delivery (CD) with the requirements of machine learning model development, testing, validation, deployment, monitoring, and re-training \cite{symeonidis2022mlops, kreuzberger2023MLOps}. 
The monitoring and re-training feedback-loop may enhance the trustworthyness of AI \cite{billeter2024mlops}. 

Together, these three overarching components address the organizational, regulatory, and methodological requirements. To tackle the complexity of securing an IT system, we adopt a layered approach, going from general IT risks to AI and RAG specific risks (Figure \ref{fig:security-layers}).

The goal of the \textbf{IT Baseline Protection layer}, is to provide general IT security.
To this end, the German \emph{Federal Office of Information Security} (BSI) developed the  \textit{IT-Grundschutz} (\textit{IT Baseline Protection} in English)) \cite{bsi_it_grundschutz}.
Similarly, the \emph{NIST Cybersecurity Framework} (CSF) \cite{nist_cybersecurity_framework} in the USA, or the international norm \emph{ISO/IEC 27001} provide standardized guidelines for general IT system protection.
These frameworks provide structured methodologies and their adoption typically progresses in breath and in depth. 
Depth refers to the level of protection. 
Profiles and tiers \cite{nist_cybersecurity_framework} allow organizations to scale their security practices based on available resources and risk tolerance.
Breath refers to the scope of a cybersecurity strategy: all components of a system need to be considered, from hardware, to data pipelines, to staff.
For example, securing a RAG at the technical level is ineffective if untrained staff disclose the same data as a result of social engineering tactics.
This comprehensive defense is exemplified by \emph{Zero Trust Architectures} (ZTA) , which shift the defense focus to users and resources (assets, services,
workflows, network accounts, etc.) \cite{NIST2020zero}. ZTA emphasizes verifying and validating users and resources before granting access, treating each service as external, and securing it accordingly. 

It should be noted, that these general frameworks must be adapted to the specific context and technologies, which can represent a considerable effort.

The \textbf{AI and LLM protection layer} establishes a secured ML system.
A practical approach is to refer to the OWASP Top 10 lists, including the \emph{OWASP Machine Learning Security Top 10} \cite{owasp_ml_top10} and the \emph{OWASP Top 10 for Large Language Model Applications} \cite{owasp_llm_top10}, which highlight critical AI security risks related to data, models, and their usage.
On a more general level, the \emph{NIST AI Risk Management Framework (RMF)} \cite{NIST2023AIRisk} promotes trustworthy and responsible AI by providing a comprehensive information security and privacy risk-management process.

Finally, the \textbf{RAG Protection layer}, addresses the task of identifying and mitigating the specific risks discussed in section \ref{section:RiskMitigations}. 
We structure the risk along four categories and, where needed, formulate questions that help identifying the risks, while table \ref{tab:risk_mitigation_table} lists the corresponding mitigation.

\textbf{General Risks:} 
The use of a RAG comes with system inherent risks, described in R0 and R1.
These risks are (partially) addressed when adopting a general security framework (see the aforementioned layers 1 and 2), while table \ref{tab:risk_mitigation_table} identifies the specific mitigations at the RAG layer.

\textbf{RAG Usage Scenarios:} 
The same attack vector can present a radically different risk depending on the scope of the RAG service. 
Public-facing applications can lead to significant reputational or legal consequences in the event of an incident.
Consequently, the protection of the (vectorized) data, and the protection against service alteration requires tailored assessment of the risks listed in Table \ref{tab:risk_mitigation_table}.
Ask: \textit{Is the RAG used privately, internally within the company, or exposed to users via a public interface?}
In addition to technical mitigations, operators must ensure that users do not place more trust in the RAG system than is justified by its accuracy, completeness, and reliability.

\textbf{External RAG Hosting:}
\textit{Where is your RAG hosted}? 
Cloud hosting and hybrid cloud hosting introduce risks R6 and R7, while on-premises does not have those forms of disclosure risks.

\textbf{Data Sources:} \textit{How is the retrieval data obtained?} Blindly scraping data from the internet, or accepting document uploads from unverified sources,introduces risks R8, and R9. 
On the other hand, if local data (for example internal documents) are used, then ask \textit{Who can manipulate the data}? If (untrusted) users can manipulate the retrieval data, then R8 applies. If only (trusted) admins have access, then R8 is not relevant.

\textbf{RAG Service Validation and Monitoring:}
The service quality may vary depending on several factors, such as the quality of the user prompt, the documents in the RAG datastore, or conversational drift over a sequence of prompts. 
Maintaining high service quality and data protection requires measures throughout all phases of the RAG service lifecycle.
The novel risks introduced by RAG and LLM systems require additional efforts during the operational phase. 
Specifically, continuous system monitoring and service validation are essential.
Existing monitoring techniques can be combined with new approaches.
For example, the sanity of the user's input prompt may be automatically verified using an additional, specifically instructed LLM. 
Similarly, the output generator's response may be passed through an LLM for validation and policy enforcement.
Instances of such systems exist, but they are often ad-hoc, and there are currently no scientifically validated solutions available.

\section{Conclusion}
RAG technology enables the integration of up-to-date, domain-specific, or proprietary knowledge into LLMs without the need for fine-tuning. 
This makes RAG an attractive technology for a variety of use cases and has led to rapid industry adoption in both internal and client-facing applications.
The goal of this research was to provide a practical and easy-to-use framework that facilitates the implementation of secure RAG systems. 
To this end, we reviewed the security risks associated with RAG and identified corresponding mitigations. 
The resulting risk and mitigation matrix offers practitioners a structured reference, which helps securing RAG systems.
Furthermore, we introduce a framework that embeds RAG-specific security considerations within existing security standards and provide implementation guidelines that emphasize the importance of a holistic approach to security.

It is important to note that implementing a secure RAG system is inherently challenging and may require significant investment. Some proposed mitigations may also impact system performance or accuracy, aspects which are not addressed in this paper.

As the technology continues to evolve, new attack vectors are likely to emerge. 
Researchers investigating new risks may benefit from our framework, as it provides an overview of the RAG attack surface.

\bibliographystyle{IEEEtran}
\bibliography{references}

\begin{thebibliography}{10}
\providecommand{\url}[1]{#1}
\csname url@samestyle\endcsname
\providecommand{\newblock}{\relax}
\providecommand{\bibinfo}[2]{#2}
\providecommand{\BIBentrySTDinterwordspacing}{\spaceskip=0pt\relax}
\providecommand{\BIBentryALTinterwordstretchfactor}{4}
\providecommand{\BIBentryALTinterwordspacing}{\spaceskip=\fontdimen2\font plus
\BIBentryALTinterwordstretchfactor\fontdimen3\font minus \fontdimen4\font\relax}
\providecommand{\BIBforeignlanguage}[2]{{%
\expandafter\ifx\csname l@#1\endcsname\relax
\typeout{** WARNING: IEEEtran.bst: No hyphenation pattern has been}%
\typeout{** loaded for the language `#1'. Using the pattern for}%
\typeout{** the default language instead.}%
\else
\language=\csname l@#1\endcsname
\fi
#2}}
\providecommand{\BIBdecl}{\relax}
\BIBdecl

\bibitem{ji-etal-2023-towards}
Z.~Ji, T.~Yu, Y.~Xu \emph{et~al.}, ``Towards mitigating {LLM} hallucination via self reflection,'' in \emph{Findings of the ACL: EMNLP 2023}.\hskip 1em plus 0.5em minus 0.4em\relax Singapore: ACL, 2023, pp. 1827--1843.

\bibitem{huang2023survey}
L.~Huang, W.~Yu, W.~Ma, W.~Zhong, Z.~Feng, H.~Wang, Q.~Chen, W.~Peng, X.~Feng, B.~Qin \emph{et~al.}, ``A survey on hallucination in large language models: Principles, taxonomy, challenges, and open questions,'' \emph{ACM Transactions on Information Systems}, 2023.

\bibitem{lewis2021retrievalaugmentedgenerationknowledgeintensivenlp}
\BIBentryALTinterwordspacing
P.~Lewis, E.~Perez, A.~Piktus, F.~Petroni, V.~Karpukhin, N.~Goyal, H.~Küttler, M.~Lewis, W.~tau Yih, T.~Rocktäschel, S.~Riedel, and D.~Kiela, ``Retrieval-augmented generation for knowledge-intensive nlp tasks,'' 2021. [Online]. Available: \url{https://arxiv.org/abs/2005.11401}
\BIBentrySTDinterwordspacing

\bibitem{zhao2024retrievalaugmentedgenerationaigeneratedcontent}
\BIBentryALTinterwordspacing
P.~Zhao, H.~Zhang, Q.~Yu, Z.~Wang, Y.~Geng, F.~Fu, L.~Yang, W.~Zhang, J.~Jiang, and B.~Cui, ``Retrieval-augmented generation for ai-generated content: A survey,'' 2024. [Online]. Available: \url{https://arxiv.org/abs/2402.19473}
\BIBentrySTDinterwordspacing

\bibitem{gao2024retrievalaugmentedgenerationlargelanguage}
\BIBentryALTinterwordspacing
Y.~Gao, Y.~Xiong, X.~Gao, K.~Jia, J.~Pan, Y.~Bi, Y.~Dai, J.~Sun, M.~Wang, and H.~Wang, ``Retrieval-augmented generation for large language models: A survey,'' 2024. [Online]. Available: \url{https://arxiv.org/abs/2312.10997}
\BIBentrySTDinterwordspacing

\bibitem{schillaci2024llmadoption}
Z.~Schillaci, ``Llm adoption trends and associated risks,'' in \emph{Large Language Models in Cybersecurity: Threats, Exposure and Mitigation}.\hskip 1em plus 0.5em minus 0.4em\relax Springer Nature Switzerland Cham, 2024, pp. 121--128.

\bibitem{gupta2023chatgpt}
M.~Gupta, C.~Akiri, K.~Aryal, E.~Parker, and L.~Praharaj, ``From chatgpt to threatgpt: Impact of generative ai in cybersecurity and privacy,'' \emph{IEEE Access}, vol.~11, pp. 80\,218--80\,245, 2023.

\bibitem{zeng2024goodbadexploringprivacy}
\BIBentryALTinterwordspacing
S.~Zeng, J.~Zhang, P.~He, Y.~Xing, Y.~Liu, H.~Xu, J.~Ren, S.~Wang, D.~Yin, Y.~Chang, and J.~Tang, ``The good and the bad: Exploring privacy issues in retrieval-augmented generation (rag),'' 2024. [Online]. Available: \url{https://arxiv.org/abs/2402.16893}
\BIBentrySTDinterwordspacing

\bibitem{haryanto2024secgenaienhancingsecuritycloudbased}
\BIBentryALTinterwordspacing
C.~Y. Haryanto, M.~H. Vu, T.~D. Nguyen, E.~Lomempow, Y.~Nurliana, and S.~Taheri, ``Secgenai: Enhancing security of cloud-based generative ai applications within australian critical technologies of national interest,'' 2024. [Online]. Available: \url{https://arxiv.org/abs/2407.01110}
\BIBentrySTDinterwordspacing

\bibitem{kucharavy2024llmcybersecuritybook}
A.~Kucharavy, O.~Plancherel, V.~Mulder, A.~Mermoud, and V.~Lenders, Eds., \emph{Large Language Models in Cybersecurity: Threats, Exposure and Mitigation}.\hskip 1em plus 0.5em minus 0.4em\relax Springer Nature Switzerland Cham, 2024.

\bibitem{owasp_llm_top10}
\BIBentryALTinterwordspacing
{Open Worldwide Application Security Project (OWASP)}, ``Owasp top 10 for large language model applications (2025),'' 2025, accessed: 2025-01-06. [Online]. Available: \url{https://genai.owasp.org/resource/owasp-top-10-for-llm-applications-2025/}
\BIBentrySTDinterwordspacing

\bibitem{mitsloan2023}
\BIBentryALTinterwordspacing
E.~M. Renieris, D.~Kiron, and S.~Mills. (2023, 6) Building robust rai programs as third-party ai tools proliferate. [Online]. Available: \url{https://sloanreview.mit.edu/projects/building-robust-rai-programs-as-third-party-ai-tools-proliferate/"}
\BIBentrySTDinterwordspacing

\bibitem{wefcybersecurity2025}
A.~Joshi, G.~Moschetta, and E.~Winslow, ``Global cybersecurity outlook 2025,'' World Economic Forum (WEF), Geneva, Switzerland, Tech. Rep., 1 2025.

\bibitem{eu_ai_regulatory_framework}
{European Commission}, ``Regulatory framework on artificial intelligence,'' \url{https://digital-strategy.ec.europa.eu/en/policies/regulatory-framework-ai}, 2024, accessed: 2025-01-06.

\bibitem{biden_executive_order_ai_2023}
J.~R.~B. Jr., ``Executive order on the safe, secure, and trustworthy development and use of artificial intelligence,'' \href{https://www.whitehouse.gov/briefing-room/presidential-actions/2023/10/30/executive-order-on-the-safe-secure-and-trustworthy-development-and-use-of-artificial-intelligence/}{White House}, October 30 2023, accessed: 2025-01-06.

\bibitem{nist_ai_rmf}
\BIBentryALTinterwordspacing
{National Institute of Standards and Technology (NIST)}, ``Artificial intelligence risk management framework (ai rmf),'' 2023, accessed: 2025-01-06. [Online]. Available: \url{https://www.nist.gov/itl/ai-risk-management-framework}
\BIBentrySTDinterwordspacing

\bibitem{greshake2023youvesignedforcompromising}
\BIBentryALTinterwordspacing
K.~Greshake, S.~Abdelnabi, S.~Mishra, C.~Endres, T.~Holz, and M.~Fritz, ``Not what you've signed up for: Compromising real-world llm-integrated applications with indirect prompt injection,'' 2023. [Online]. Available: \url{https://arxiv.org/abs/2302.12173}
\BIBentrySTDinterwordspacing

\bibitem{huang2024transferableembeddinginversionattack}
\BIBentryALTinterwordspacing
Y.-H. Huang, Y.~Tsai, H.~Hsiao, H.-Y. Lin, and S.-D. Lin, ``Transferable embedding inversion attack: Uncovering privacy risks in text embeddings without model queries,'' 2024. [Online]. Available: \url{https://arxiv.org/abs/2406.10280}
\BIBentrySTDinterwordspacing

\bibitem{zou2024poisonedragknowledgecorruptionattacks}
\BIBentryALTinterwordspacing
W.~Zou, R.~Geng, B.~Wang, and J.~Jia, ``Poisonedrag: Knowledge corruption attacks to retrieval-augmented generation of large language models,'' 2024. [Online]. Available: \url{https://arxiv.org/abs/2402.07867}
\BIBentrySTDinterwordspacing

\bibitem{xu2024knowledgeconflictsllmssurvey}
\BIBentryALTinterwordspacing
R.~Xu, Z.~Qi, Z.~Guo, C.~Wang, H.~Wang, Y.~Zhang, and W.~Xu, ``Knowledge conflicts for llms: A survey,'' 2024. [Online]. Available: \url{https://arxiv.org/abs/2403.08319}
\BIBentrySTDinterwordspacing

\bibitem{zhang2024hijackraghijackingattacksretrievalaugmented}
\BIBentryALTinterwordspacing
Y.~Zhang, Q.~Li, T.~Du, X.~Zhang, X.~Zhao, Z.~Feng, and J.~Yin, ``Hijackrag: Hijacking attacks against retrieval-augmented large language models,'' 2024. [Online]. Available: \url{https://arxiv.org/abs/2410.22832}
\BIBentrySTDinterwordspacing

\bibitem{ablon2017zero}
L.~Ablon and A.~Bogart, \emph{Zero days, thousands of nights: The life and times of zero-day vulnerabilities and their exploits}.\hskip 1em plus 0.5em minus 0.4em\relax Rand Corporation, 2017.

\bibitem{chang2023surveyevaluationlargelanguage}
\BIBentryALTinterwordspacing
Y.~Chang, X.~Wang, J.~Wang, Y.~Wu, L.~Yang, K.~Zhu, H.~Chen, X.~Yi, C.~Wang, Y.~Wang, W.~Ye, Y.~Zhang, Y.~Chang, P.~S. Yu, Q.~Yang, and X.~Xie, ``A survey on evaluation of large language models,'' 2023. [Online]. Available: \url{https://arxiv.org/abs/2307.03109}
\BIBentrySTDinterwordspacing

\bibitem{guo2023evaluatinglargelanguagemodels}
\BIBentryALTinterwordspacing
Z.~Guo, R.~Jin, C.~Liu, Y.~Huang, D.~Shi, Supryadi, L.~Yu, Y.~Liu, J.~Li, B.~Xiong, and D.~Xiong, ``Evaluating large language models: A comprehensive survey,'' 2023. [Online]. Available: \url{https://arxiv.org/abs/2310.19736}
\BIBentrySTDinterwordspacing

\bibitem{morris2023textembeddingsrevealalmost}
\BIBentryALTinterwordspacing
J.~X. Morris, V.~Kuleshov, V.~Shmatikov, and A.~M. Rush, ``Text embeddings reveal (almost) as much as text,'' 2023. [Online]. Available: \url{https://arxiv.org/abs/2310.06816}
\BIBentrySTDinterwordspacing

\bibitem{anderson2024dataretrievaldatabasemembership}
\BIBentryALTinterwordspacing
M.~Anderson, G.~Amit, and A.~Goldsteen, ``Is my data in your retrieval database? membership inference attacks against retrieval augmented generation,'' 2024. [Online]. Available: \url{https://arxiv.org/abs/2405.20446}
\BIBentrySTDinterwordspacing

\bibitem{yang2023review}
J.~Yang, Y.-L. Chen, L.~Y. Por, and C.~S. Ku, ``A systematic literature review of information security in chatbots,'' \emph{MDPI Applied Sciences}, vol.~13, no.~11, 2023.

\bibitem{jiang2024ragthief}
\BIBentryALTinterwordspacing
C.~Jiang, X.~Pan, G.~Hong, C.~Bao, and M.~Yang, ``Rag-thief: Scalable extraction of private data from retrieval-augmented generation applications with agent-based attacks,'' 2024. [Online]. Available: \url{https://arxiv.org/abs/2411.14110}
\BIBentrySTDinterwordspacing

\bibitem{duan2024privacyriskincontextlearning}
\BIBentryALTinterwordspacing
H.~Duan, A.~Dziedzic, M.~Yaghini, N.~Papernot, and F.~Boenisch, ``On the privacy risk of in-context learning,'' 2024. [Online]. Available: \url{https://arxiv.org/abs/2411.10512}
\BIBentrySTDinterwordspacing

\bibitem{chu2024reconstructpreviousconversationscomprehensively}
\BIBentryALTinterwordspacing
J.~Chu, Z.~Sha, M.~Backes, and Y.~Zhang, ``Reconstruct your previous conversations! comprehensively investigating privacy leakage risks in conversations with gpt models,'' 2024. [Online]. Available: \url{https://arxiv.org/abs/2402.02987}
\BIBentrySTDinterwordspacing

\bibitem{chen2024blackboxopinionmanipulationattacks}
\BIBentryALTinterwordspacing
Z.~Chen, J.~Liu, H.~Liu, Q.~Cheng, F.~Zhang, W.~Lu, and X.~Liu, ``Black-box opinion manipulation attacks to retrieval-augmented generation of large language models,'' 2024. [Online]. Available: \url{https://arxiv.org/abs/2407.13757}
\BIBentrySTDinterwordspacing

\bibitem{wang2024poisonedlangchainjailbreakllms}
\BIBentryALTinterwordspacing
Z.~Wang, J.~Liu, S.~Zhang, and Y.~Yang, ``Poisoned langchain: Jailbreak llms by langchain,'' 2024. [Online]. Available: \url{https://arxiv.org/abs/2406.18122}
\BIBentrySTDinterwordspacing

\bibitem{deng2024pandorajailbreakgptsretrieval}
\BIBentryALTinterwordspacing
G.~Deng, Y.~Liu, K.~Wang, Y.~Li, T.~Zhang, and Y.~Liu, ``Pandora: Jailbreak gpts by retrieval augmented generation poisoning,'' 2024. [Online]. Available: \url{https://arxiv.org/abs/2402.08416}
\BIBentrySTDinterwordspacing

\bibitem{stam2020data}
A.~Stam and B.~Kleiner, ``Data anonymization: legal, ethical, and strategic considerations,'' \emph{FORS Guide No. 11, Version 1.0. Lausanne: Swiss Centre of Expertise in the Social Sciences FORS}, 2020.

\bibitem{raghunathan2021synthetic}
T.~E. Raghunathan, ``Synthetic data,'' \emph{Annual Review of Statistics and Its Application}, vol.~8, pp. 129--140, 2021.

\bibitem{zeng2024mitigatingprivacyissuesretrievalaugmented}
\BIBentryALTinterwordspacing
S.~Zeng, J.~Zhang, P.~He, J.~Ren, T.~Zheng, H.~Lu, H.~Xu, H.~Liu, Y.~Xing, and J.~Tang, ``Mitigating the privacy issues in retrieval-augmented generation (rag) via pure synthetic data,'' 2024. [Online]. Available: \url{https://arxiv.org/abs/2406.14773}
\BIBentrySTDinterwordspacing

\bibitem{namer2024retrieval}
\BIBentryALTinterwordspacing
A.~Namer, R.~Lagerstr{\"o}m, G.~Balakrishnan, and B.~Maltzman, ``Retrieval-augmented generation (rag) classification and access control,'' \emph{Technical Disclosure Commons, Defensive Publications Series}, September 2024. [Online]. Available: \url{https://www.tdcommons.org/dpubs_series/}
\BIBentrySTDinterwordspacing

\bibitem{xu2025mixtureofinstructionsaligninglargelanguage}
\BIBentryALTinterwordspacing
B.~Xu, S.~Wu, K.~Liu, and L.~Hu, ``Mixture-of-instructions: Aligning large language models via mixture prompting,'' 2025. [Online]. Available: \url{https://arxiv.org/abs/2404.18410}
\BIBentrySTDinterwordspacing

\bibitem{asai2023selfraglearningretrievegenerate}
\BIBentryALTinterwordspacing
A.~Asai, Z.~Wu, Y.~Wang, A.~Sil, and H.~Hajishirzi, ``Self-rag: Learning to retrieve, generate, and critique through self-reflection,'' 2023. [Online]. Available: \url{https://arxiv.org/abs/2310.11511}
\BIBentrySTDinterwordspacing

\bibitem{edge2024localglobalgraphrag}
\BIBentryALTinterwordspacing
D.~Edge, H.~Trinh, N.~Cheng, J.~Bradley, A.~Chao, A.~Mody, S.~Truitt, and J.~Larson, ``From local to global: A graph rag approach to query-focused summarization,'' 2024. [Online]. Available: \url{https://arxiv.org/abs/2404.16130}
\BIBentrySTDinterwordspacing

\bibitem{besta2024multiheadragsolvingmultiaspect}
\BIBentryALTinterwordspacing
M.~Besta, A.~Kubicek, R.~Niggli, R.~Gerstenberger, L.~Weitzendorf, M.~Chi, P.~Iff, J.~Gajda, P.~Nyczyk, J.~Müller, H.~Niewiadomski, M.~Chrapek, M.~Podstawski, and T.~Hoefler, ``Multi-head rag: Solving multi-aspect problems with llms,'' 2024. [Online]. Available: \url{https://arxiv.org/abs/2406.05085}
\BIBentrySTDinterwordspacing

\bibitem{geminiteam2024gemini}
\BIBentryALTinterwordspacing
G.~T. et~al., ``Gemini 1.5: Unlocking multimodal understanding across millions of tokens of context,'' 2024. [Online]. Available: \url{https://arxiv.org/abs/2403.05530}
\BIBentrySTDinterwordspacing

\bibitem{kamps2007Eval}
J.~Kamps, M.~Lalmas, and J.~Pehcevski, ``Evaluating relevant in context: document retrieval with a twist,'' in \emph{Proceedings of the 30th Annual International ACM SIGIR Conference on Research and Development in Information Retrieval}.\hskip 1em plus 0.5em minus 0.4em\relax New York, NY, USA: Association for Computing Machinery, 2007, p. 749–750.

\bibitem{papineni-etal-2002-bleu}
K.~Papineni, S.~Roukos, T.~Ward, and W.-J. Zhu, ``{B}leu: a method for automatic evaluation of machine translation,'' in \emph{Proceedings of the 40th Annual Meeting of the Association for Computational Linguistics}.\hskip 1em plus 0.5em minus 0.4em\relax Association for Computational Linguistics, Jul. 2002, pp. 311--318.

\bibitem{lin2004rouge}
C.-Y. Lin, ``Rouge: A package for automatic evaluation of summaries,'' in \emph{Text summarization branches out}.\hskip 1em plus 0.5em minus 0.4em\relax Association for Computational Linguistics, 2004, pp. 74--81.

\bibitem{zhang2020bertscoreevaluatingtextgeneration}
\BIBentryALTinterwordspacing
T.~Zhang, V.~Kishore, F.~Wu, K.~Q. Weinberger, and Y.~Artzi, ``Bertscore: Evaluating text generation with bert,'' 2020. [Online]. Available: \url{https://arxiv.org/abs/1904.09675}
\BIBentrySTDinterwordspacing

\bibitem{sander2021exam}
\BIBentryALTinterwordspacing
D.~P. Sander and L.~Dietz, ``Exam: How to evaluate retrieve-and-generate systems for users who do not (yet) know what they want,'' in \emph{DESIRES}, 2021, pp. 136--146. [Online]. Available: \url{https://ceur-ws.org/Vol-2950/paper-16.pdf}
\BIBentrySTDinterwordspacing

\bibitem{ragas2023}
ExplodingGradients, ``Ragas: Evaluation framework for retrieval-augmented generation,'' \url{https://github.com/explodinggradients/ragas}, 2023, accessed: 2025-01-08.

\bibitem{saadfalcon2024aresautomatedevaluationframework}
\BIBentryALTinterwordspacing
J.~Saad-Falcon, O.~Khattab, C.~Potts, and M.~Zaharia, ``Ares: An automated evaluation framework for retrieval-augmented generation systems,'' 2024. [Online]. Available: \url{https://arxiv.org/abs/2311.09476}
\BIBentrySTDinterwordspacing

\bibitem{powerOfNoiseRAG}
F.~Cuconasu, G.~Trappolini, F.~Siciliano, S.~Filice, C.~Campagnano, Y.~Maarek, N.~Tonellotto, and F.~Silvestri, ``The power of noise: Redefining retrieval for rag systems,'' in \emph{The Power of Noise: Redefining Retrieval for RAG Systems}.\hskip 1em plus 0.5em minus 0.4em\relax New York, NY, USA: Association for Computing Machinery, 2024.

\bibitem{rerankerSchemeNLU}
C.~Su, R.~Gupta, S.~Ananthakrishnan, and S.~Matsoukas, ``A re-ranker scheme for integrating large scale nlu models,'' in \emph{2018 IEEE Spoken Language Technology Workshop (SLT)}, 2018, pp. 670--676.

\bibitem{alessio2024improving}
M.~Alessio, G.~Faggioli, N.~Ferro, F.~M. Nardini, R.~Perego \emph{et~al.}, ``Improving rag systems via sentence clustering and reordering,'' in \emph{RAG@ SIGIR 2024 workshop: The Information Retrieval’s Role in RAG Systems, ACM}, 2024, pp. 1--10.

\bibitem{euGDPR}
{European Parliament, Council of the European Union}, ``{General Data Protection Regulation (GDPR)},'' 2016, \url{https://gdpr-info.eu/}.

\bibitem{NIST2024RiskManagementFramework}
\BIBentryALTinterwordspacing
``{NIST Risk Management Framework (RMF)}.'' [Online]. Available: \url{https://csrc.nist.gov/projects/risk-management}
\BIBentrySTDinterwordspacing

\bibitem{euAIact}
{European Parliament, Council of the European Union}, ``{The EU Artificial Intelligence Act (EU AI Act)},'' 2024, \url{https://eur-lex.europa.eu/eli/reg/2024/1689/oj}.

\bibitem{janssen2020data}
M.~Janssen, P.~Brous, E.~Estevez, L.~S. Barbosa, and T.~Janowski, ``Data governance: Organizing data for trustworthy artificial intelligence,'' \emph{Government information quarterly}, vol.~37, no.~3, p. 101493, 2020.

\bibitem{blivznak2024systematic}
K.~Bli{\v{z}}n{\'a}k, M.~Munk, and A.~Pilkov{\'a}, ``A systematic review of recent literature on data governance (2017-2023),'' \emph{IEEE Access}, 2024.

\bibitem{symeonidis2022mlops}
G.~Symeonidis, E.~Nerantzis, A.~Kazakis, and G.~Papakostas, ``Mlops - definitions, tools and challenges,'' in \emph{2022 IEEE 12th Annual Computing and Communication Workshop and Conference (CCWC)}.\hskip 1em plus 0.5em minus 0.4em\relax IEEE, 2022, pp. 0453--0460.

\bibitem{kreuzberger2023MLOps}
D.~Kreuzberger, N.~K{\"u}hl, and S.~Hirschl, ``Machine learning operations (mlops): Overview, definition, and architecture,'' \emph{IEEE access}, vol.~11, pp. 31\,866--31\,879, 2023.

\bibitem{billeter2024mlops}
Y.~Billeter, P.~Denzel, R.~Chavarriaga, O.~Forster, F.~Schilling, S.~Brunner, C.~Frischknecht-Gruber, M.~Reif, and J.~Weng, ``Mlops as enabler of trustworthy ai,'' in \emph{2024 11th IEEE Swiss Conference on Data Science (SDS)}.\hskip 1em plus 0.5em minus 0.4em\relax IEEE, 2024, pp. 37--40.

\bibitem{bsi_it_grundschutz}
{Federal Office for Information Security (BSI)}, ``It-grundschutz: Standards and certification for information security,'' \url{https://www.bsi.bund.de/DE/Themen/Unternehmen-und-Organisationen/Standards-und-Zertifizierung/IT-Grundschutz/it-grundschutz_node.html}, 2025, accessed: 2025-01-08.

\bibitem{nist_cybersecurity_framework}
{National Institute of Standards and Technology (NIST)}, ``Cybersecurity framework,'' \url{https://www.nist.gov/cyberframework}, 2025, accessed: 2025-01-08.

\bibitem{NIST2020zero}
S.~Rose, S.~Mitchell, and S.~Connelly, ``{NIST Zero Trust Architecture},'' 2020.

\bibitem{owasp_ml_top10}
\BIBentryALTinterwordspacing
{Open Worldwide Application Security Project (OWASP)}, ``Owasp machine learning security top ten,'' 2023, accessed: 2025-01-06. [Online]. Available: \url{https://owasp.org/www-project-machine-learning-security-top-10/}
\BIBentrySTDinterwordspacing

\bibitem{NIST2023AIRisk}
``{NIST Artificial Intelligence Risk Management Framework (AI RMF 1.0)},'' 2023.

\end{thebibliography}

\end{document}